# Chemical-Grafting of Graphene Oxide Quantum Dots for Enhanced Electrochemical Analysis of Aromatics

Albina Mikhraliieva[1], Volodymyr Zaitsev[1,2]*, Oleg Tkachenko[3,4], Michael Nazarkovsky[1] and Edilson V. Benvenutti[4]

[1] Department of Chemistry, Pontifical Catholic University of Rio de Janeiro, Marquês de São Vicente, 225, 22451-900, Rio de Janeiro, Brasil

[2] National University of Kyiv-Mohyla Academy, 2 Skovorody vul., Kyiv, Ukraine 04070

[3] Materials Chemistry Department, V. N.Karazin Kharkiv National University, 4 Svoboda Square, Kharkiv, 61022, Ukraine

[4] Institute of Chemistry, UFRGS, PO Box 15003, CEP, Porto Alegre, RS, 91501-970, Brazil

*E-mail: vnzaitsev@puc-rio.br

**Abstract**

Because of high surface area and combination of various functional groups, graphene oxide (GO) is currently one of the most actively studying materials for electroanalytical applications. Self-supported GO is not practical enough to be utilized individually and thus is commonly integrated with different supporting carriers. Having large lateral size GO can only wrap the particles of the support and thus, can significantly reduce the surface area of porous materials. To reach synergy form high surface area and polyfunctional nature of GO, and a rigid structure of porous support, the lateral size of GO must be essentially decreased. Recently reported graphene oxide quantum dots (GOQDs) can fulfill this task. We report successful preparation of mesoporous silica supported GOQDs exhibiting strong luminescent band with the maximum at 404 nm. From $N_2$ adsorption isotherms, it was demonstrated that the surface area of the resulted hybrid material was augmented. Raman spectrum of $SiO_2$-GOQDs shows two distinct peaks at 1585 $cm^{-1}$(G-peak) and 1372 $cm^{-1}$ (D-peak) indicating presence ordered basal plane of graphene with aromatic $sp^2$ domains, and disordered oxygen-containing structure. Covalent immobilization of GOQDs on aminosilica via such oxygen-containing fragments was proved by FTIR, MAS $^{13}C$ MNR, and XPS spectroscopy. $SiO_2$-GOQDs was used as a modifier of carbon paste electrode for differential pulse voltammetry determination of four environmental: sulfamethoxazole, trimethoprim, diethylstilbestrol (DES), and estriol (EST). The modified electrodes demonstrated an essential decrease in LOD for EST (220%) and DES (760%), which was explained by significant $\pi$-$\pi$ stacking interaction between GOQDs and aromatic system of the analytes.

## Introduction

Graphene oxide (GO) is one of the most commonly used carbon modifiers for preparation of various hybrid materials [1, 2]. Graphene oxide belongs to the class of 2D-nano objects (nanosheets) having up to several nanometers of thickness with the lateral size >$10^4$ nm. Macromolecule of GO has a huge surface area (up to 2630 $m^2$ $g^{-1}$ [3]) and contains a basal plane of graphene with many oxygen-containing defects. Thus, it can interact with various molecules via non-covalent bonding including electrostatic, hydrogen and dative bonds, π-π stacking and dispersion forces [4, 5]. Because of its high surface area and polyfunctional nature GO has a great potential to be used for preconcentration of organic compounds containing aromatic rings and electronegative functional groups [6–9].

Self-supported 2D materials are not practical enough to be utilized individually. Due to strong π-π stacking and hydrophobic interaction between the graphene layers GO nanosheets can easily agglomerate and thus, drastically reduce their surface area. Therefore, GO is commonly integrated with different supporting carriers, such as mesoporous silica gels [10, 11], magnetic nanoparticles [12], carbon nanotubes, inorganic oxides, such as $TiO_2$, $Fe_2O_3$, MgO, etc., [13]. Since the lateral size of GO particles is about 10-100 μm which is 2-20 times bigger than in common silica particles, GO can only wrap silica particles (see, for example [10, 14]). If we take into account that at least 90% of the surface area of mesoporous silicas allocated in the pores, it becomes obvious that immobilization of flat non-porous nanosheet onto a surface of porous support instead of increasing, can significantly reduce the overall surface area of a hybrid material due to pore blocking. As a result, the adsorption capacity of $SiO_2$@GO hybrid materials towards analytes can be much lower than individual GO and even $SiO_2$. For example, $SiO_2$@GO demonstrates ten times lower total adsorption capacity to Cu and Pb than silica-based adsorbent [14]. Unfortunately, in many recent publications, this effect was ignored that thus hybrid $SiO_2$@GO materials have not shown their full potential [1, 15–17].

To reach synergy form high surface area and polyfunctional nature of GO, and a rigid structure of the porous support, the lateral size of GO must be essentially decreased. A lot of progress has been achieved in this direction, to tuned properties of GO by downsizing its particles up to several nanometer scales [18–20]. New particles received by the downsizing of GO were signified to graphene oxide quantum dots (GOQDs) [21, 22]. Immobilized GOQDs demonstrate improved properties as adsorbents for HPLC [23], photonic/electronic devices [24], as carriers for photo-thermal and redox-responsive release of medications [25], as sensors in the electrochemical analysis [26]. In the latter case, smaller nanoparticles have better electrochemical properties [20].

Because of the miniaturization of electronics in recent years, the interest in electroanalytical methods has been growing, particularly in the rapid inexpensive determination of environmentally important contaminants in-field with portable instruments. The technical development associated with electrochemical sensors is an efficient, low-cost, fast-response and easy-to-operate alternative compared to spectroscopic or chromatographic ones. Besides, electrochemical devices have the advantage of portability and miniaturization [27]. Various materials have been used as working electrode in such devices: conductive glasses [28], screen-printed [29], glassy carbon [30], ceramic carbon [31], and carbon paste [32]. To increase the electroactive area and facilitate the kinetics of the electrode/solution interface and thus sensitivity, the electrodes have commonly modified with metal/metal oxide nanoparticles [33, 34], carbon nanotubes [35], graphene [36] or GO [37], or with hybrid silica-based materials [38]. For example, it was found that the addition of mesoporous silica (SBA-15) to carbon paste electrode (CPE) can considerably enhance its sensitivity toward diethylstilbestrol [39], as much as immobilization of graphene on glassy carbon electrode [40]. Application of GOQDs as individual modifier as well as a part of hybrid material is under intensive electrochemical studies [29, 41]. For instance, the possibility of simultaneous determination of dopamine and epinephrine using gold nanocrystals capped with graphene quantum dots in a silica network was demonstrated recently [42].

Among biologically active compounds, which are inevitably discharged into the environment antibiotics and hormones require special attention. The residue of the antibiotics in the environment has resulted in bacterial resistance, which could seriously affect human health and ecological balance [43]. Environmental estrogens belong to a group of endocrine disruptors (ED) and can cause cancerous tumors, birth defects, and other developmental disorders of the endocrine system even at low concentrations [44–46]. Hence, it is important to develop analytical methods for time- and cost-effective monitoring of ED in various media. It seems that electrochemical sensors can resolve this demand [47].

The idea of the current research was to integrate supporting and size-selective properties of mesoporous silica network, and electroactive properties of GOQD in order to obtain high surface area working electrode with an enhanced affinity towards chemicals having aromatic or heterocycle fragments. As model compounds, we selected next endocrine disruptors: sulfamethoxazole (SMZ) and trimethoprim (TMP) (antibiotics), diethylstilbestrol (DES) and estriol (EST) (hormones). Selected medications have various polar functional groups as well as aromatic and heterocyclic fragments, allowing their multiple-point interaction with GOQDs surface, Fig. 1.

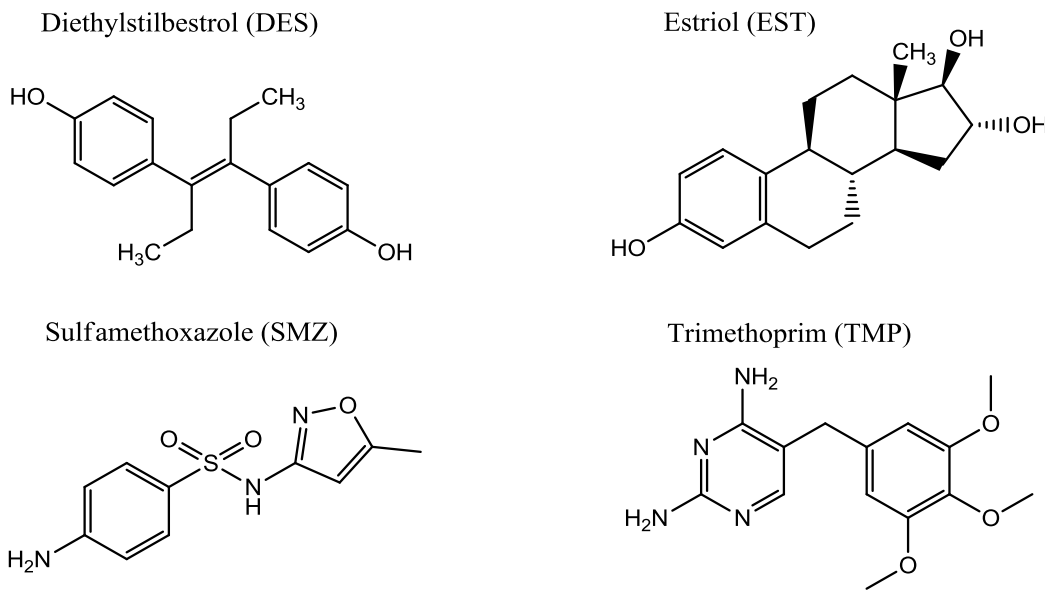

Fig. 1 Chemical formulas of medications studied in this research

This interaction can increase the sensitivity of the electrode towards analytes and improve the selectivity of the analysis. With this aim in view, hybrid $SiO_2$-GOQDs material was prepared and characterized by MAS NMR $^{13}C$, XPS, Raman and FTIR spectroscopies, electron microscopy (SEM) and solid-phase photoluminescence spectroscopy. To prevent possible side-reactions on electrocatalytically active Mn-sites [48–50] special attention was given on the preparation of Mn-free GOQD.

## Materials and methods

*Chemicals and reagents*

Silica gel with 4 nm average pore size and 63-200 μm particle size distribution was purchased from Merck, (3-Aminopropyl)triethoxysilane (APTES, ≥98%), toluene anhydrous, N,N′-dicyclohexylcarbodiimide (DCC, 99%), potassium permanganate (KMnO4, ≥99.0%), hydrogen peroxide solution ($H_2O_2$, 30 % (w/w)) were purchased from Sigma–Aldrich. Ninhydrin (99%) and sulfuric acid (95-98%) were provided by QHEMIS (Brazil), graphite and dimethylformamide (99.8%) – from Synth (Brazil). Diethylstilbestrol, estriol (both 97%), sulfamethoxazole (SMZ, 98%) and trimethoprim (TMP, 98%) were from Sigma-Aldrich. Britton-Robinson buffer (0.04 mol $L^{-1}$) was prepared from the following reagents: boric acid (Neon, 99.5%), acetic acid (Dinamica, 99.7%), phosphoric acid (Vetec, 85%), hydrochloric acid (Merck, 37%), sodium hydroxide (Vetec, 97%), sodium nitrate (Quimica Moderna, >99%). After distillation with calcium hydride (Sigma-Aldrich), toluene was kept in a dark bottle with 3 Å molecular sieves (4-8 mesh, Sigma-Aldrich). The aqueous solutions were prepared using ultra-pure water from PURELAB Classic, Elga, UK).

*Characterization techniques*

The solution pH was measured using a PHS-3E pH-meter with BioTrode (Hamilton, USA) ion-selective electrode. The electrical conductivities of suspensions were measured using HI 8633 conductivity meter (HANNA instruments, UK). The concentrations of manganese ions in solutions were determined by Optima 7300 DV Inductively Coupled Plasma Optical Emission Spectrometry (Perkin Elmer), and in ZEEnit 60 Atomic Absorption Spectrometer with a transversely-heated graphite furnace (Analytik Jena AG, Germany). The specific surface area and pore size of silica gels were determined from nitrogen adsorption/desorption isotherms on Tristar II 3020 Kr (Micromeritics, USA). Elemental analysis (CHN) of the samples was done on PE-2400 elemental analyzer (Perkin Elmer, USA) and loading of immobilized groups was calculated from Eq.1.

$$C_L (mmol\ g^{-1}) = \frac{10 \cdot P_e}{A_r \cdot n_e},\qquad \text{Eq.1}$$

where $P_e$ is the content of element, %; $A_r$ is the element atomic mass; $n_e$ is the number of atoms of the element in the grafted fragment.

Fourier transform infrared spectroscopy (FTIR) spectra were recorded in the region 4000-400 cm$^{-1}$ on FTLA-2000 spectrometer (Thermo Scientific Nicolet). Solid-state $^{13}$C and $^{29}$Si CP/MAS NMR of the samples were obtained on Agilent Technologies DD2 500/54A (Agilent, USA) - 100.6 MHz ($^{13}$C). UV–Vis absorption spectra were measured on a Cary 100 (Agilent, USA). The composition of hybrid materials was determined from X-ray photoelectron spectroscopy (XPS), using a K-Alpha X-ray photoelectron spectrometer (Thermo Fisher Scientific, UK) equipped with a hemispherical electron analyzer and an aluminum anode X-ray source (Kα = 1486.6 eV) at the energy resolution of ~1 eV. For organic analytes, octane-water distribution coefficients (P) were calculated using software ChemDraw Ultra 12.

The concentration of immobilized functional groups ($C_L$) was calculated from Eq. 2.

$$C_L (mmol\ g^{-1}) = \frac{1000 \cdot at(N)}{at(Si) \cdot M(SiO_2)}\qquad \text{Eq. 2}$$

where $at(Si)$ and $at(N)$ are the atomic percentage of Si and N in the hybrid, respectively; $M(SiO_2)$ is SiO$_2$ molar mass.

The morphology of the samples was studied on a low-vacuum JSM-6490LV scanning electron microscope (SEM) (JEOL, Japan). Differential pulse voltammograms (DPVs) were recorded on IviumStat potentiostat/galvanostat (Ivium Technologies, The Netherlands) with a conventional three-electrode cell. Modified carbon paste electrode (CPE, working electrode), platinum wire (the auxiliary electrode) and Ag/AgCl (reference electrode) were mounted in the cell. For DPV the following parameters were used: pulse amplitude of 50 mV, pulse time of 50 ms, step potential of 1 mV and scan rate of 10 mV s$^{-1}$. From each voltammogram the background was subtracted, as commonly recommended [51]. The limit of detection (LOD) and limit of quantification (LOQ) were calculated as next: LOD = 3$S_b$/b, and LOQ = 10$S_b$/b where **$S_b$** is the standard deviation of the blank (n = 10) and **b** is the slope of the calibration curve.

*Synthesis of graphene oxide quantum dots*

GOQDs were prepared in one-step ultrasonic synthesis [52]. In brief, 200 mL of a mixture of concentrated $H_2SO_4$ and $H_3PO_4$ (9:1 v/v) was added to graphite powder (1.5 g) in a 500-mL round-bottom flask equipped with a mechanical stir bar. Then $KMnO_4$ (9 g) was added slowly at room temperature under stirring and after the flask was heated to 50 °C and kept at this temperature for 12 h. The obtained lite-pink mixture (Fig. S1a) was cooled down, poured slowly onto ice (400 mL) with 30% $H_2O_2$ (20 mL) giving orange suspension, Fig S1b. The solid was separated by centrifugation and washed with water. To remove manganese impurities, the precipitate was immersed into HCl (2%) then separated from the solution by centrifugation. The procedure was repeated until negative results on Mn in solution with ICP-OES. Finally, 40 mg of freeze-dried film (Fig S1d) was sonicated in 50 mL of DMF for about 2 h. Emitting green-blue light in UV irradiation dispersion was subjected to further immobilization.

*Preparation of $SiO_2$-GOQDs hybrid material*

GOQDs were covalently immobilized on silica gel surface via silica-immobilized aminosilane and carboxylic groups of GOQDs, as described elsewhere [7]. Typically, activated in $HNO_3$ and dried at 500 °C silica gel (10 g) was suspended in 100 mL of dry toluene and 3 mL of APTES was added under constant stirring. The reaction mixture was refluxed for 10 h, filtrated, washed in a Soxhlet apparatus with toluene for 24 h and finally dried in vacuum at 120 °C for 7 h. The resulted aminosilica ($SiO_2$-$NH_2$), (3 g) was added to the suspension of GOQDs in 50 mL of DMF and 40 mg of DCC were added under stirring. The suspension was heated at 85 °C for 60 h with periodic sonification for 30 min. Solid-phase was separated by decantation, washed with DMF, methanol and water under ultrasonic treatment (5 min). Finally, the precipitate was dried at 120 °C for 8 h to obtain about 3 g of $SiO_2$-GOQDs. Chemical analysis of $SiO_2$-$NH_2$ and $SiO_2$-GOQDs revealed augmented carbon content in the samples - from 2.89 to 3.22% (Table S1), while the percentage of nitrogen did not change. This indicated that about 3.3 mg of GOQDs has been immobilized per one gram of $SiO_2$-GOQDs, which constitute 25% from initially loaded for synthesis (13 mg/g).

*Preparation of CPE modified electrode*

Typically, 8 mg of $SiO_2$-GOQDs and 12 mg of graphite powder were carefully mixed in an agate mortar with addition of mineral oil (5 mg). The prepared homogeneous paste was placed in a Teflon cavity (1 mm in depth and 2 mm in diameter), covered with a platinum disk fused to a glass tube with copper wire as an electric conductor. The manufactured electrode was denoted as CPE/Si/GOQD. Unmodified CPE electrode, fabricated in the same way without $SiO_2$-GOQDs was used for the comparative analyses.

## Results and discussion

*Synthesis approach*

Hybrid $SiO_2$-GOQDs materials can be obtained by adsorption and further chemical anchoring of the preliminarily prepared GOQDs on the support surface [4, 25, 53]. As an alternative way is incomplete pyrolysis of small organic compounds trapped inside of the adsorbed pores [54]. In the latter case, porous matrix can better confine the size and shape of the resulted GOQDs [55, 56]. However, preparation of $SiO_2$-GOQDs inside the host porous system can have an undesirable effect, since it can, as mentioned above, essentially reduce the specific surface area of the resulted hybrid material. For example, SBA15-GQDs nanocomposite prepared by incomplete pyrolysis of pyrene adsorbed in SBA15 pores has only 26 $m^2$ $g^{-1}$ while the specific surface area of the pristine host was 719 $m^2$ $g^{-1}$ [55]. Therefore, the first scheme has been selected and GOQDs were prepared from graphite in one-step ultrasonic synthesis with further immobilization of appropriate nanoparticles in the porous system of mesoporous silica. The ability of GOQDs to infiltrate into silica pores have been confirmed recently [26, 57]. Samples of GO were prepared from graphite powder by oxidation with $KMnO_4$ in the mixture of concentrated $H_2SO_4$ and $H_3PO_4$, because it was demonstrated that the introduction of $H_3PO_4$ to the acid mixture improved safety of the process and led to the successful increase in the reaction yield [58].

It was important to obtain metal-free nanocomposite [55] since even traces of electrocatalytically active metal ions such as Mn can corrupt properties of $SiO_2$-GOQDs electrodes [48, 59, 60]. To ensure Mn-free composition of GO samples, the concentration of Mn ions in washing solutions was monitored. The results presented in Fig. S2 demonstrates that for removal of $Mn^{2+}$ ions impurities from GO by water are less sufficient than washing with 2% HCl. This fact can be explained by good adsorption properties of GO towards metal ions [1, 61].

Covalent immobilization of GOQDs on silica surface was performed via well-established procedure of $SiO_2$-$NH_2$ acylation in anhydrous solvent (DMF) by carboxylic functional groups of GOQDs in the presence of DCC [10, 14]. Excess of GOQDs was separated from the final product by multiple decantations of the precipitate alternating with the ultrasonic treatment of $SiO_2$-GOQDs in DMF. Finally obtained grey product demonstrates greenish luminescence under irradiation with UV light at 365 nm, Fig. 1d.

*Characterization of $SiO_2$-GOQDs*

**Nitrogen adsorption/desorption measurements**

From Fig. 2, it can be seen that samples $SiO_2$-GOQDs have the identical character of $N_2$ adsorption/desorption isotherm that pristine $SiO_2$-$NH_2$. This suggests that immobilization of GOQDs does not change the porous structure of silica support having typical Type IV isotherm with a distinct hysteresis loop of H1 within the $p/p_0$ range of 0.4–1.0, indicating mesopore presence. At high relative pressure, the saturation of the isotherms is clearly observed and this feature indicates the complete filling of the mesopores and the absence of macropores. The very similar isotherm profile and surface area values evidence that the addition of GOQDs does not produce significant textural changes. However, from slightly higher average pore volume and pore size for $SiO_2$-$NH_2$ and larger surface area for $SiO_2$-GOQDs (Table 1) incorporation of GOQDs to the porous structure of hybrid material can be assumed.

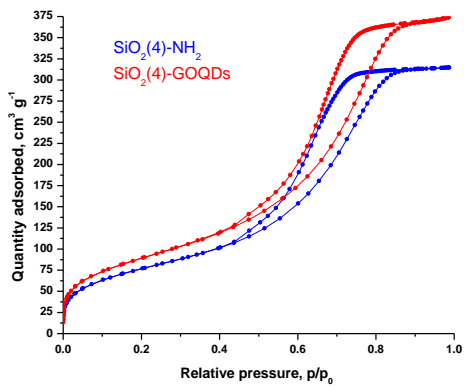

Fig. 2 $N_2$ adsorption/desorption isotherms of $SiO_2$-$NH_2$ (blue curve) and $SiO_2$-GOQDs (red curve)

The maintenance of the textural properties is an excellent result regarding that the material will be applied as an electrochemical sensor. The presence of graphene domains provides electrical conductance and at the same time, the material maintains the porosity that allows the diffusion of the electroactive species and provides high electroactive area [38].

The BET surface area of modified material increases by about 15%. This increase was interpreted considering a contribution from the graphene oxide that presents high surface area. This contribution can be seen from the profile of the isotherms. After the graphene oxide addition, an increase in the amount of adsorbed nitrogen occurs, even in low relative pressures (P/P0 < 0.2), which is related to the micropores, probably contribution of the graphene sheets [3].

Table 1. The textural characteristics of the obtained materials.

| Material | Average pore size, nm | $S_{BET}$, $m^2 g^{-1}$ |
|---|---|---|
| $SiO_2$-$NH_2$ | 5.4 | 278 |
| $SiO_2$-GOQDs | 5.2 | 324 |

**SEM**

The morphology of GO and $SiO_2$-GOQDs was investigated by SEM technique. Freeze-dried GO demonstrates a closely packed lamellar texture reflecting its multilayered microstructure, (Fig. 3a inset). With the exfoliation of graphite oxide into GO, the edges of the GO sheets become crumpled, folded, and closely restacked. The GO sheet has a rough surface of the film (Fig. 3a). SEM image of SiO$_2$-GOQDs obviously demonstrates that silica particles have not wrapped by GO and thus porous structure of the hybrid SiO$_2$-GOQDs composite is open. It is in agreement with the conclusions made from N$_2$-adsorption isotherms above.

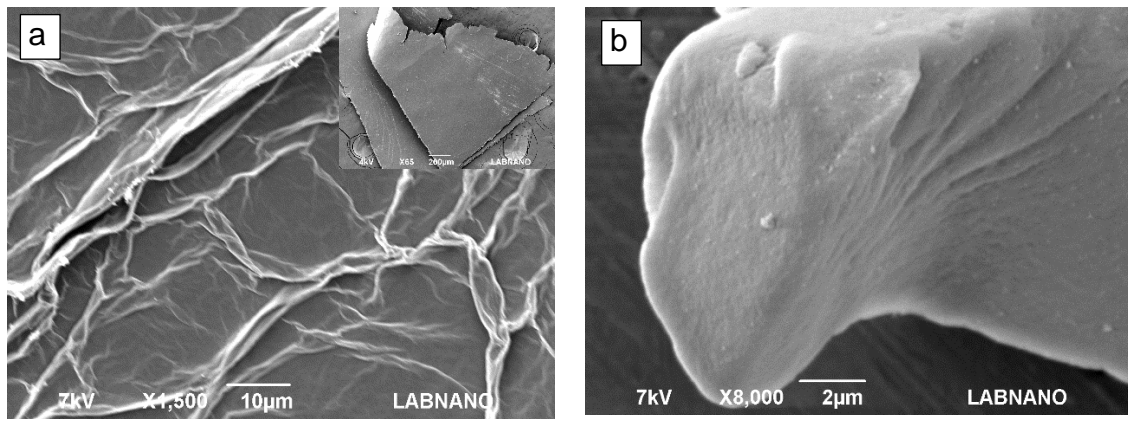

Fig. 3 SEM images of GO(a) and SiO$_2$-GOQDs (b)

*Fluorescence spectroscopy*

Photoluminescent spectra of SiO$_2$-GOQDs serve as positive proof of GOQDs immobilization on SiO$_2$ surface. It is known that neither aminosilica, nor GO exhibit photoluminescence, but SiO$_2$-GOQDs exhibit strong luminescent band with the maximum at 404 nm (Fig. 4), likewise GOQDs in water suspension [24] and other hybrid silica-based materials with loaded GOQDs [21, 26].

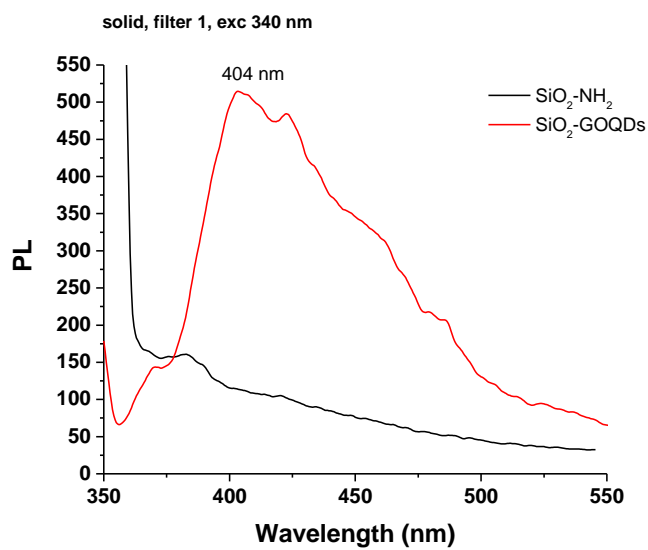

Fig. 4 Photoluminescent spectra of solid SiO$_2$-NH$_2$ (—) and SiO$_2$-GOQDs (—) under excitation at 340 nm.

*Raman spectroscopy*

To investigate the successful synthesis of GOQDs and immobilization of the nanoparticles obtained on silica surface, the Raman spectra of GOQDs and SiO$_2$-GOQDs were recorded. From Fig. 5 it is evident that the Raman spectrum for GOQDs shows two distinct peaks at 1585 cm$^{-1}$ (G-peak) and 1372 cm$^{-1}$ (D-peak).

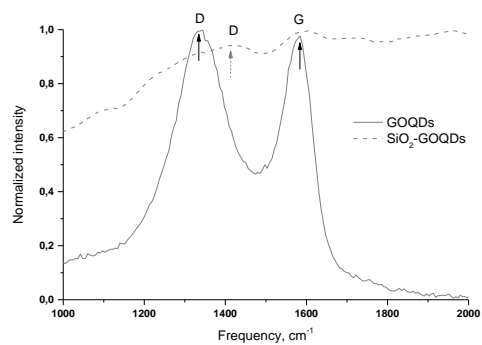

Fig. 5 Raman spectra of GOQDs (—) and SiO$_2$-GOQDs (- -).

The D peak is an indication of the disordered structure of graphene in GOQDs due to oxidation. The G peak from the ordered basal plane of graphene is observed at 1585 cm$^{-1}$. The ratio between the intensities of characteristic bands ($I_D/I_G$) gives an indication of the functional group insertion [62]. For experimental samples $I_D/I_G < 1$ presumably indicating a decrease in the fraction of aromatic sp$^2$ domains in GOQDs with increasing the number of detection oxygen-containing sites [63]. Further similar evidence for oxygen-containing groups in GOQDs was received from FTIR, NMR MAS and XPS of the SiO$_2$-GOQDs hybrid material. Despite strong fluorescence background and low carbon content, we were able to record Raman spectrum from SiO$_2$-GOQDs and it confirms immobilization of GOQDs, Fig. 5. Ordered G band was detected at about 1585 cm$^{-1}$ that matches the position of G band of individual GOQDs indicating negligible interaction between silica scaffold and basal plane of immobilized GOQDs. Contrary, the position of D-band in hybrid material is shifted to 1412 cm$^{-1}$ Fig. 5, suggesting anchoring of GOQDs via oxygen-containing sites. It is obvious that not all such sites of GOQDs react with immobilized aminosilane fragments in SiO$_2$-NH$_2$. Therefore D-band in Raman spectra of SiO$_2$-GOQDs is proliferated, Fig. 5.

*FTIR characterization*

The FTIR spectra of SiO$_2$-GOQDs, as well as Raman spectra, suggest the presence of oxygen-containing sites in GOQDs, including C(O)O–H ($v_{O-H}$ at 3390 cm$^{-1}$), CO–H ($v_{O-H}$ at 3250 cm$^{-1}$) and carboxyl ($v_{OC=O}$ at 1730 cm$^{-1}$) moieties [6], Fig. 6.

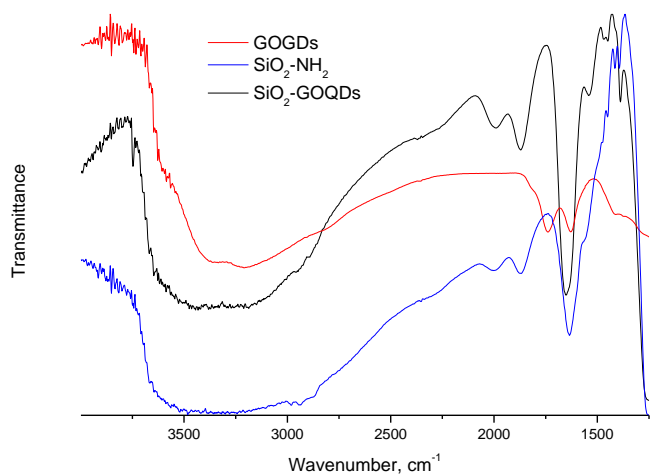

Fig. 6 FTIR spectra of GOQDs, SiO$_2$-NH$_2$, SiO$_2$-GOQDs

Apart of silica gel matrix, the pristine SiO$_2$-NH$_2$ shows several bands at about 2950 cm$^{-1}$ ($v_{CH3}$ and $v_{CH2}$), and at peaks at 1560, 1475, 1450 1420 and 1390 cm$^{-1}$ that correspond to stretching vibrations of propylamine chain. FTIR spectrum of GOQDs essentially changed on immobilization. Particularly, stretching vibration of the carboxylic group at 1730 cm$^{-1}$ disappeared, and bands at 1650 cm$^{-1}$ and 1574 cm$^{-1}$ corresponding to stretching vibration of C=O and bending vibrations of N–H in NHC(O) fragment of immobilized moiety developed instead, Fig. 6. From these observations, covalent immobilization of GOQDs via carboxyl fragments of the nanoparticle and aminogroup of the silane can be assumed [6, 23].

*MAS 13C NMR spectroscopy of SiO$_2$-GOQDs.*

The SiO$_2$-GOQDs have low (less than four percent) total carbon content, therefore $^{13}$C MAS NMR spectrum of this nanocomposite is very noisy, Fig. 7.

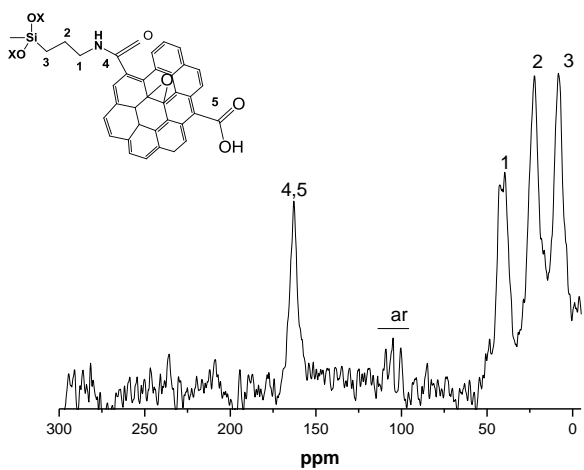

Fig. 7 CP/MAS $^{13}$C NMR spectrum of SiO$_2$-GOQDs

Nevertheless, we were able to record the spectra and identify the signals. Intensive peaks at 10, 25, and 40 ppm were correspondingly assigned to Si-$\underline{C}_\alpha$H$_2$, CH$_2$-$\underline{C}_\beta$H$_2$ and $\underline{C}_\gamma$H$_2$-NH- in immobilized moiety, Fig. 7. A signal at 165 ppm was identified as carbonyl fragment of GOQDs [64, 65], while series of signals at 100-115 ppm were assigned to sp$^2$ carbons in the basial plane of graphene [66]. Consequently, $^{13}$C NMR spectrum of SiO$_2$-GOQDs as well as FTIR and Raman demonstrate an essential fraction of oxygen-containing fragments in GOQDs and prove immobilization of the GOQDs in the porous scaffold of silica.

*XPS spectroscopy*

XPS spectra of SiO$_2$-GOQDs further confirmed the successful immobilization of nanoparticles. As demonstrated, SiO$_2$-GOQDs is Mn-free material, since its XPS does not contain Mn 2p peaks at 641.3 and 653.2 eV [67], Fig. 8a.

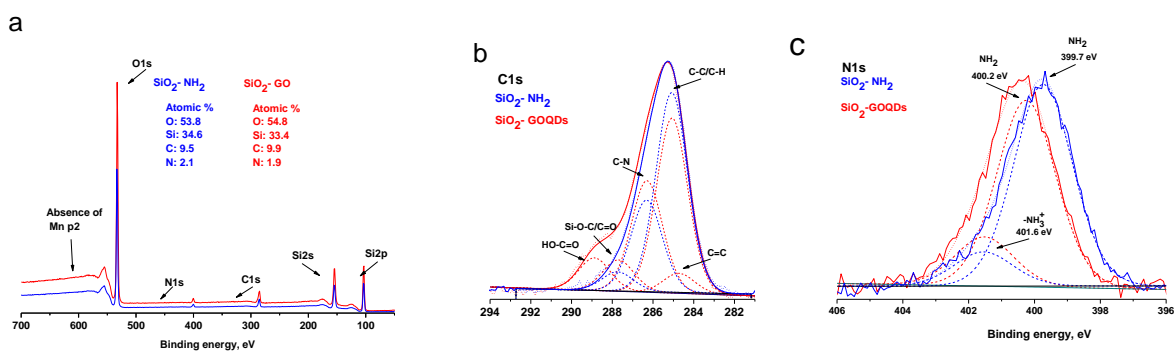

Fig. 8 Survey (a) and fitted XPS spectra of the C1s (b) and N1s (c) of SiO$_2$-GOQDs and pristine SiO$_2$-NH$_2$ with deconvoluted data (d) regions corresponding to SiO$_2$-NH$_2$ and SiO$_2$-GOQDs.

The high-resolution XPS spectra demonstrate the essential difference between C1s bands for SiO$_2$-NH$_2$ and SiO$_2$-GOQDs, Fig. 8b. The C1s band from SiO$_2$-NH$_2$ can be deconvoluted into three components attributed to C-C, C-N and C-O bonds in SiO$_2$-immobilized aminopropyl fragments, Table 2. The relative intensity of C-C and C-N peaks in the spectrum is about 3:1, which correlates with the composition of the immobilized fragment. Also, about 6% of carbon atoms in SiO$_2$-NH$_2$ bonded with oxygen, which lets us assume the occurrence of incomplete hydrolysis of ethoxy groups of aminosilane in the immobilization process, as it is demonstrated in Fig. 7(inset). The high-resolution XPS C1s spectra of SiO$_2$-GOQDs among the signals from SiO$_2$-NH$_2$ shows additional peaks (Fig. 8b), attributed to C=C and C=O bonds in O-C=O or N-C=O fragments (Table 2), indicating successful immobilization of GOQDs [14].

Table 2. XPS analysis of SiO$_2$-GOQDs and pristine SiO$_2$-NH$_2$

| SiO$_2$-NH$_2$ | SiO$_2$-GOQDs |
|---|---|
| C1s ||

| Type | C-H, C-C | C-N | C-O | C=C | CH, C-C | C-N | C-O | OCO |
|---|---|---|---|---|---|---|---|---|
| eV | 285.1 | 286.3 | 287.7 | 284.7 | 285.1 | 286.2 | 287.7 | 288.9 |
| % | 69 | 25 | 6 | 5 | 47 | 31 | 8 | 9 |
| N1S | | | | | | | | |
| | $NH_2$ | $NH_3^+$ | | NH | | $NH_3^+$ | | |
| eV | 399.7 | 401.6 | | 400.2 | | 401.6 | | |
| % | 86 | 14 | | 79 | | 21 | | |
| $C_L$ = 1.01 mmol g$^{-1}$ | | | | $C_L$ = 0.95 mmol g$^{-1}$ | | | | |

Fig. 8c illustrates the high-resolution N1s XPS spectra of the synthesized hybrids. Band of pristine SiO$_2$-NH$_2$ consists of components attributed to neutral and protonated primary amines. On GOQDs immobilization fraction of H-bonded amines noticeably increased, while another fraction of amine fragments was transformed to amide, Table 2. This effect reflects a peculiarity of the surface reaction of the immobilized amine with nanoparticles containing several carboxyl fragments. It should be readily apparent that only few carboxyl groups of GOQDs can acylate immobilized amines due to steric restrictions. Others will be ionized and will protonate the remaining amines, as it is illustrated in Fig. 9.

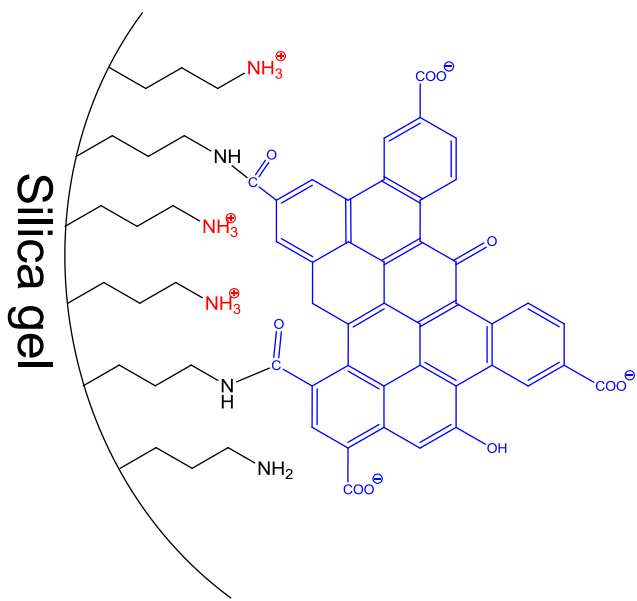

Fig. 9 Schematic structure of surface layer in SiO$_2$-GOQDs

*Electrochemical properties of the carbon paste electrode modified with SiO$_2$-GOQDs*

Carbon paste electrode (CPE) was selected for modification because it is cheap, can be reproducibly fabricated in any laboratory and SiO$_2$-GOQDs can be easily integrated into the electrode. Two antibiotics and two hormones were selected for investigation, namely: sulfamethoxazole and trimethoprim, diethylstilbestrol and estrogen estriol, Fig. 1. Diethylstilbestrol is the first synthetic estrogen that had been extensively used in the treatment of estrogen-deficiency disorders. Nowadays, DES is often illegally used as growth-promoter for cattle, sheep and poultry in many countries [68]. Although it had been prohibited as a growth promoter for years, these estrogens are still found in the river [69], fish [70], milk [71], and meat [8].

All selected analytes have aromatic rings that can form π-π stacking complexes with GOQDs, but they also have various polar fragments, with can weaken or enhance such interaction and thus change the sensitivity of the electrochemical analysis. First, the CPE electrode modified with $SiO_2$-GOQDs was tested in differential pulse voltammetry (DVP) oxidation of a mixture, containing sulfamethoxazole and trimethoprim (5:1). The results were compared with those, obtained on the CPE electrode without any additives.

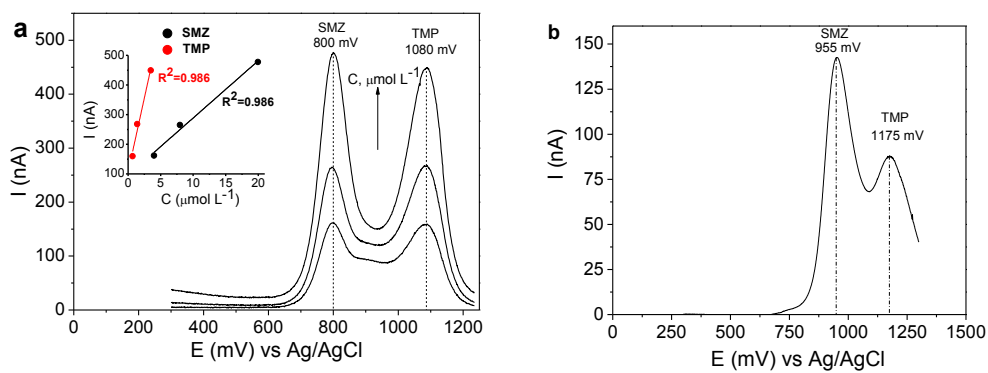

Fig. 10. DPV analysis of sulfamethoxazole (SMZ) and trimethoprim (TMP) on modified with $SiO_2$-GOQDs (a) and bulk (b) carbon paste electrodes. The analytes were present in a mixture (5:1) with the next concentration of SMZ: a) 4.0, 8.0 and 20 µmol L$^{-1}$; b) 4.0 µmol L$^{-1}$. Supporting electrolyte: 0.04 mol L$^{-1}$ of BRbs (pH 5.8), 0.5 mol L$^{-1}$ of $NaNO_3$.

From the data presented in Fig. 10 it is amply clear that electrode modified with $SiO_2$-GOQDs has higher response to TMP (87% of enhancement) and only 14% to SMZ. Further, the oxidation peaks are shifted to lower potentials for both components but differently (-155 mV for SMZ and -95 mV for TMP). This effect contributes to the difference in anodic peaks of analytes to 280 mV making determination of SMZ and TMP in their simultaneous presence more reliable. Improved sensitivity and better-separated peaks point to a benefit of $SiO_2$-GOQDs on electrochemical analysis of SMZ and TMP by DVP.

The significant improvement of the electrocatalytic performance of the modified electrode was also observed in the DVP analysis of both selected hormones. In DES analysis the modified electrode demonstrated 7.6-fold enhancement of sensibility and by 2.2-fold in EST analysis compared to the bare electrode, Fig. 11. Similarly, to the results of the antibiotics analysis, the oxidation peaks of both hormones are shifted to lower potentials for SiO$_2$-GOQDs modified electrode in comparison with the neat one, attesting better interaction of the analytes with active centers of the electrode.

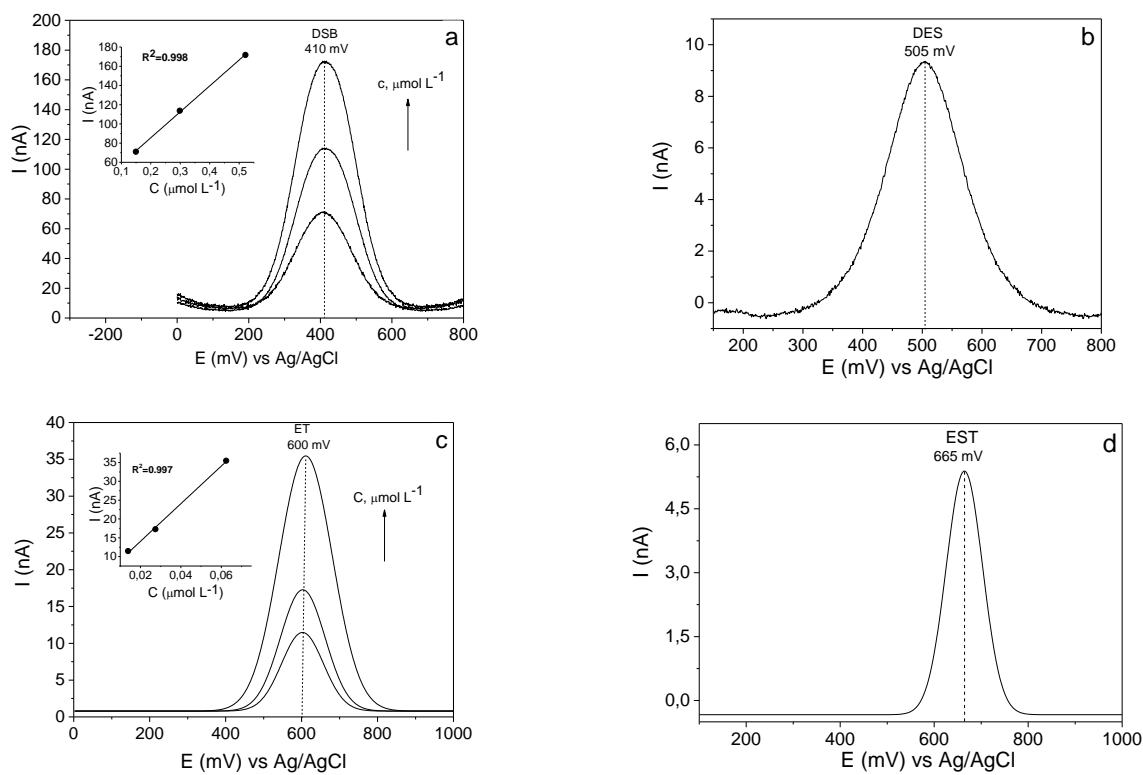

Fig. 11 DPV analysis of DES (a,b) and EST (c,d) on modified with SiO$_2$-GOQDs (a,c) and bulk (b,d) carbon paste electrodes. Concentration of the analytes: DES - 0.15, 0.30 and 0.52 µmol L$^{-1}$, EST - 0.014, 0.027, 0.062 µmol L$^{-1}$). Supporting electrolyte: 0.04 mol L$^{-1}$ of BRbs (pH 5.8), 0.5 mol L$^{-1}$ of NaNO$_3$.

The modified electrodes demonstrated linear signal response *vs.* concentration of the analytes according to the equations that are presented in Table S2 along with LOD and LOQ for the current modified electrode and other electrodes reported earlier. From the data presented it can be seen that CPE modified with SiO$_2$-GOQDs evinces the lowest value of LOD for EST and DES. These results can be explained considering the textural characteristics and chemical nature of SiO$_2$-GOQDs. As demonstrated earlier, the integrated of mesoporous silica to graphite paste electrode can enhance its electrochemical response due to the enlargement of the electroactive sites on the electrode surface [38]. Immobilization of GOQDs improves the affinity of the resulted modified electrode to the analytes that can interact with immobilized electroactive fragments, such as basial aromatic domains in GOQDs. Indeed, from the analysis of the results presented in the Table S2, electrode modified with SiO$_2$-GOQDs has much higher sensitivity towards more lipophilic EST (500 nA L µmol$^{-1}$) and DES (270 nA L µmol$^{-1}$) then to SMZ (19 nA L µmol$^{-1}$) and TMP (100 nA L µmol$^{-1}$), which are more polar (Fig. S3). But only higher lipophilicity of the hormones versus the antibiotics cannot explain the correlation observer for SiO$_2$-GOQDs modified electrode. Indeed, enhancement in the electrochemical response of modified electrode was augmented in the next series of analytes: SMZ (14%) < TMP (87%) < EST (220%) < DES (760%). This correlation corresponds to the ability of the analytes to form $\pi$-$\pi$ stacking complexes. Particularly, DES is the only compound among investigated, containing two conjugated benzene rings. EST possesses similar molecular geometry but with a single aromatic ring and thus it is less able to form stable $\pi$-$\pi$ stacking complexes. Identically to TMP, SMZ has one aromatic and heterocycle ring, but it has negatively charged sulfamide fragment (Fig S3), which repels from negatively charged GOQDs surface (Fig. 9). Thus, the detection of SMZ was least sensitive among the four substances.

## Conclusions

The results demonstrate a great potential of GOQDs for application in electrochemistry. In contrast to GO, nanoparticles of GOQDs can infiltrate in the porous system of the support and thus maintain high surface area of the electrode. CPE modified with $SiO_2$-GOQDs demonstrated enhanced sensitivity towards DES and EST and less effective towards TMP and SMZ. This fact was explained by significant $\pi$-$\pi$ stacking interaction between immobilized GOQDs and selected hormones. Oxidation peaks for all analytes were shifted to lower potentials for about 100 – 150 mV, demonstrating better interaction between the analytes and active centers of the electrode.

## Acknowledgments

Volodymyr Zaitsev is grateful for the financial support received from Conselho Nacional de Desenvolvimento Científico e Tecnológico (CNPq) (grants 306992/2018-3 and 438450/2018-3) and Fundação Carlos Chagas Filho de Amparo à Pesquisa do Estado do Rio de Janeiro (FAPERJ) (grants E-26/010.000978/2019, E-26/210.547/2019). Mikhraliieva is grateful to CNPq (154820/2015-6) and (FAPERJ) (E-26/200.612/2018) for the conceded fellowships. Nazarkovsky thanks to Coordenação de Aperfeiçoamento de Pessoal de Nível Superior (CAPES) for receiving funds (grant № 2013037-31005012005P5 – PNPD-PUC Rio) to carry out the research. We also appreciate the technical support received from Brazilian Nanotechnology National Laboratory (LNNano) in XPS and Institute of Chemistry at UFRGS in CP/MAS 13C NMR. Authors thank LabNano (Brazilian Center for Research in Physics, CBPF, Brazil) for continued assistance in the microscopy studies. Also, we thank Dr. Omar Pandoli (PUC-Rio) for Raman measurements.

## Conflict of interests

The authors declare that they have no conflict of interest.

**Graphical abstract**

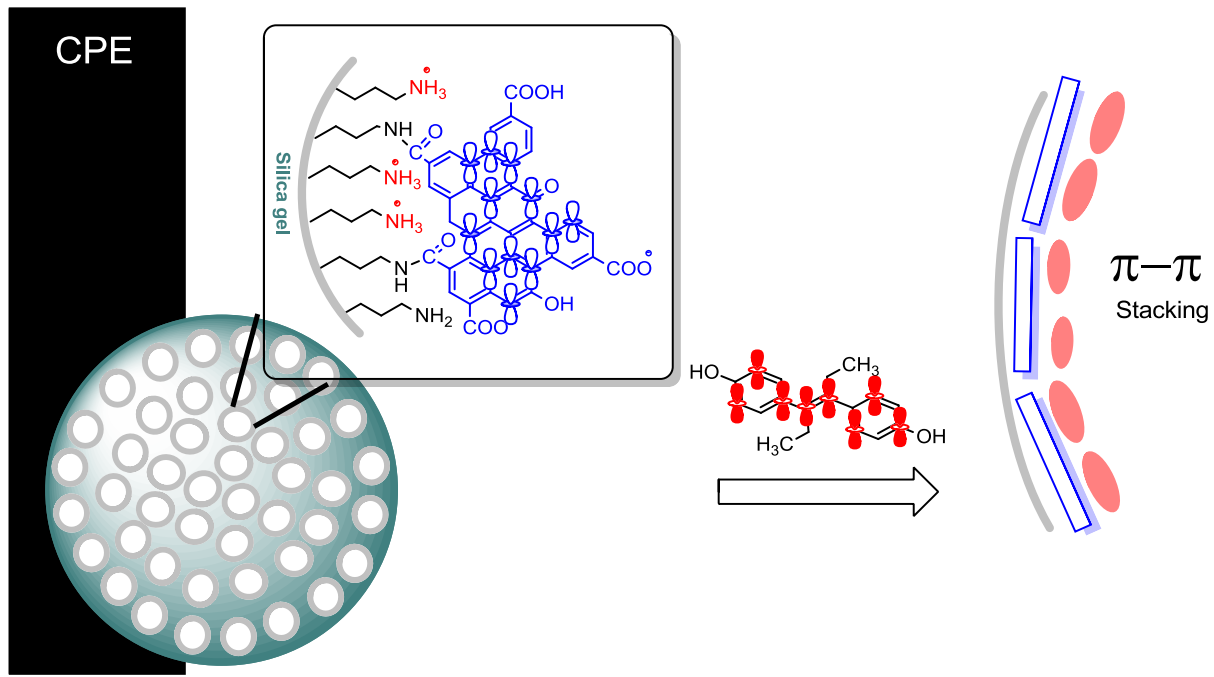

**Supplementary information**

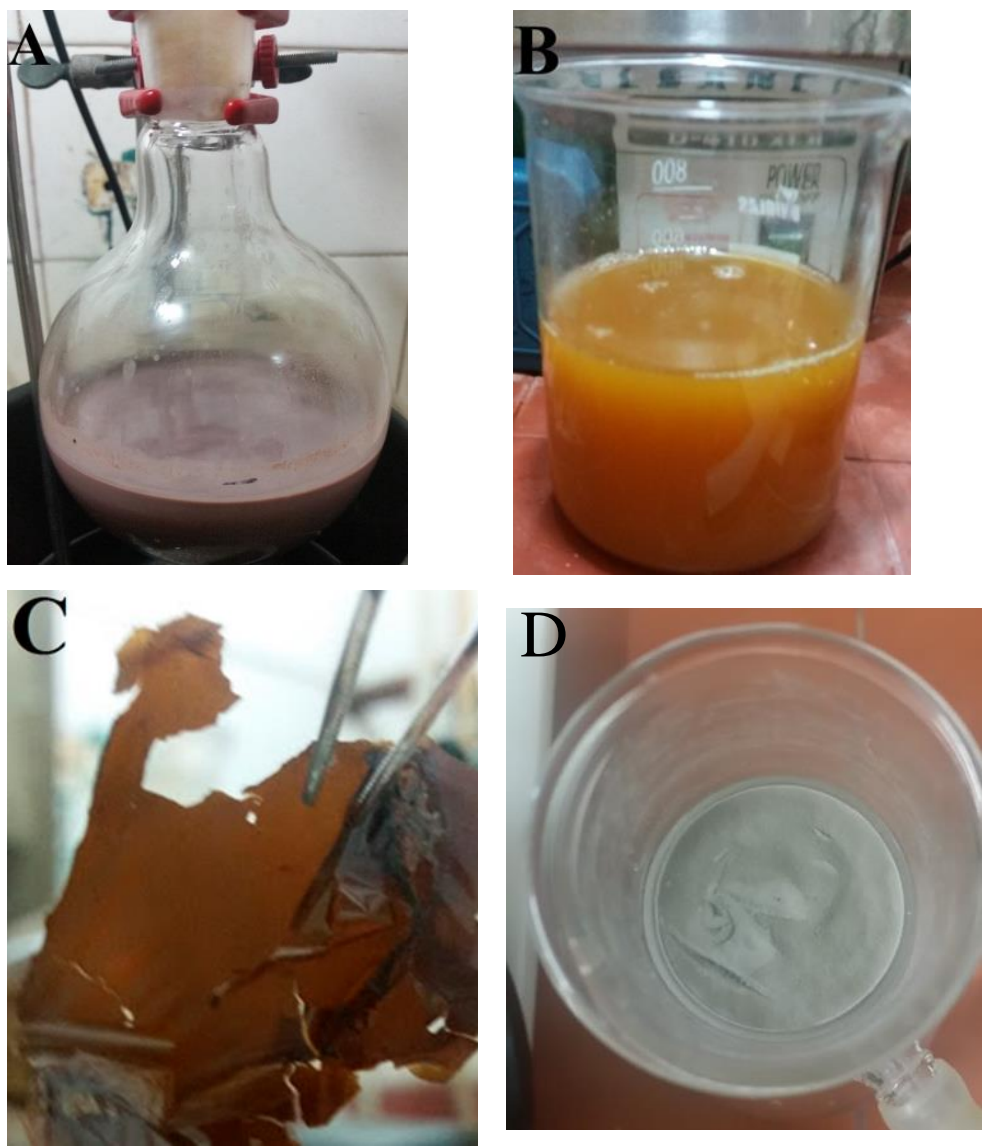

Fig. S1. Digital photos of the reaction mixture after heating during oxidation of graphite (A), after addition of ice and $H_2O_2$ (B), dry graphene oxide (C) and composite $SiO_2$-GOQDs (D)

Table S1 - The results of elemental analysis

| Material | N,% | C,% | H,% | $C_L^N$, mmol/g |
|---|---|---|---|---|
| $SiO_2$-$NH_2$ | 1.05 | 2.89 | 0.84 | 0.75 |
| $SiO_2$-GOQDs | 1.00 | 3.22 | 0.86 | |

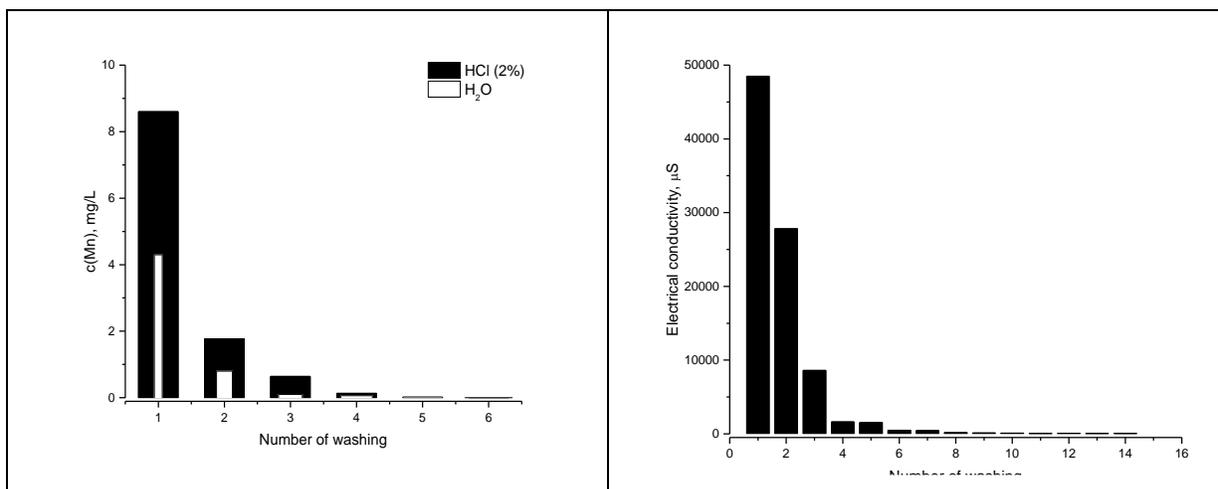

Fig. S2 Effect of washing on c(Mn) in supernatants (a) and monitoring of electrical conductivity with water washing (b)

Table S2 Analytical characteristics of the different electrodes modified with carbogenic nanomaterials in determination of EST, DES, SMZ and TMP

| Material | Analyte | Linear range, µmol$^{-1}$ | LOD, µmol$^{-1}$ | LOQ. µmol$^{-1}$ | Equation (y(nA)vsC(µmol$^{-1}$)) | Sensitivity, nA L µmol$^{-1}$ | Ref. |
|---|---|---|---|---|---|---|---|
| CPE/SiO$_2$-GOQDs | EST | 0.01-0.6 | 0.009 | 0.029 | y=500*C+4.1 | 500 | This work |
| GCE/CNB/AgNPs | | 0.2-3 | 0.16 | 0.5 | n.d. | 131 | [72] |
| GCE/rGO/AgNPs | | 0.1-3 | 0.021 | | y=590*C+590 | 590 | [73] |
| GCE/rGO/SbNPs | | 0.2-1.4 | 0.0005 | | y=2.13*C+0.57 | 2.13 | [74] |
| GCE/Pt/CNTs | | 0.5-15 | 0.62 | | y=790*C+22040 | 790 | [75] |
| CPE/SiO$_2$-GOQDs | DES | 0.15-0.5 | 0.18 | 0.6 | y=270*C+32 | 270 | This work |
| CPE | | 0.1-15 | 0.01 | 0.03 | n.d. | n.d. | [76] |
| CPE/SBA-15 | | 0.0075-0.3 | 0.003 | | n.d. | n.d. | [39] |
| GCE | | 2-100 | 0.08 | | y=337*C+264 | 337 | [77] |
| GCE/GO/CS | | 0.015-30 | 0.003 | 0.01 | y=−69.1*C−1.83 | 69 | [40] |
| GCE/rGO/CD | | 0.01-13 | 0.004 | | n.d. | n.d. | [78] |
| CPE/SiO$_2$-GOQDs | SMZ | 4-20 | 0.46 | 1.53 | y=19*C+95.90 | 19 | This work |
| CPE/MCM-41 | | 98-327 | 3.1 | | y=2.43*C+11352 | 2.43 | [38] |
| CPE/CNTs | | 1.4-119 | 0.4 | 1.33 | n.d. | 24.1 | [32] |
| SPE/rGO | | 0.5–50 | 0.04 | 0.13 | n.d. | n.d. | [79] |
| MIP/BDD | | 0.1-100 | 0.024 | 0.080 | y=314.22*C+7.17 | 314 | [43] |
| CPE/SiO$_2$-GOQDs | TMP | 0.7-3.5 | 0.191 | 0.63 | y=100*C+106 | 100 | This work |
| SPE/CNTs/PBnc | | 0.1-10 | 0.06 | 0.2 | y=108.31*C+70.7 | 108 | [80] |
| CPE/CNTs/SbNPs | | 0.1–0.7 | 0.031 | 0.1 | y=0.37*C+30 | 0.37 | [81] |

Electrodes: CPE – carbon paste electrode; GCE - glassy carbon electrode; MIP - molecularly imprinted polymer; SPE - screen-printed electrode.

Modifiers: CNB – carbon black nanoballs; AgNPs - silver nanoparticles; SbNPs - antimony nanoparticles; AuNPs - gold nanoparticles; rGO - reduced graphene oxide; CoPc - cobalt phtolocyanine; CS - chitosan; CD - β-cyclodextrin; CNTs – carbon nanotubes; BDD - boron-doped diamond; PBnc - Prussian blue nanocubes.

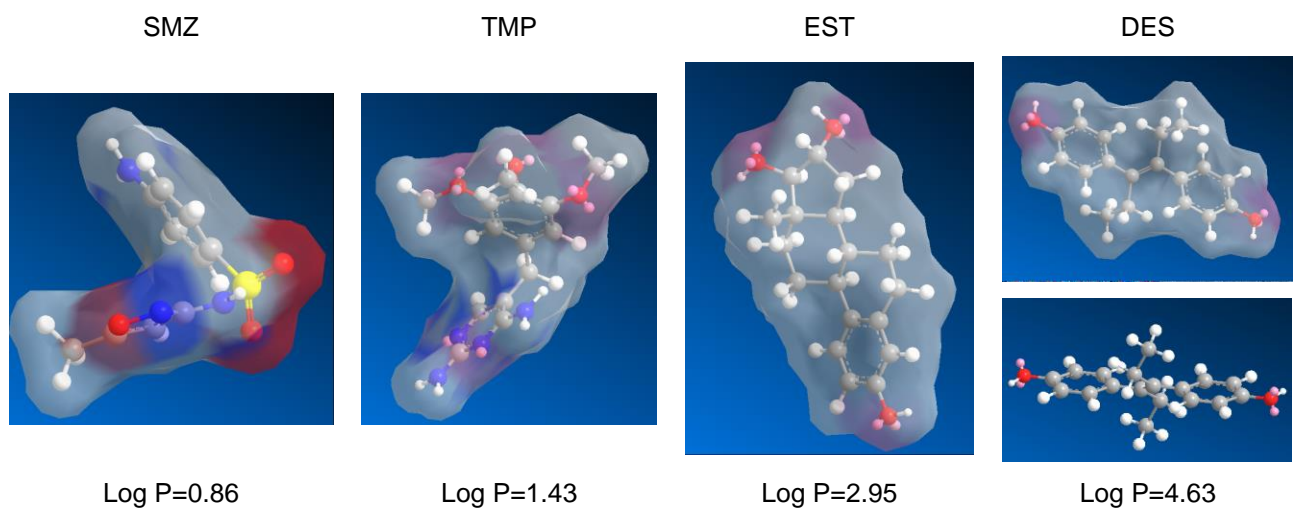

Fig. S 3. Polarity of the selected analytes calculated by software ChemDraw Ultra 12.0